\documentclass[fleqn]{aa}
\usepackage[fleqn]{amsmath}
\usepackage{amssymb}
\usepackage{graphicx}
\usepackage{caption}
\usepackage{subcaption}
\usepackage{grffile} 
\usepackage{natbib}
\usepackage{txfonts}
\usepackage{color}
\usepackage{multirow}
\usepackage{tabu}
\usepackage{caption}
\usepackage{subcaption}

\usepackage[]{hyperref}
\hypersetup{
breaklinks = {true},
colorlinks = {false},
pdfpagemode = {None}, 
pdfborder = {0 0 1},
pdftitle = {Testing semi-analytic models using galaxy-galaxy(-galaxy) lensing observations},
pdfsubject = {study of models of galaxy evolution with galaxy-galaxy(-galaxy) lensing aperture statistics},
pdfauthor = {Hananeh Saghiha, Patrick Simon, Peter Schneider, and Stefan Hilbert},
pdfkeywords = {gravitational lensing: weak -- large-scale structure of the Universe -- cosmology: observations -- galaxies: formation -- galaxies: evolution -- methods: numerical}
}

\newcommand{\TCB}[1]{\textcolor{black}{#1}}

\newcommand{\vect}[1]{\boldsymbol{#1}}

\newcommand{\ev}[1]{\left\langle{#1}\right\rangle}
\newcommand{\bev}[1]{\bigl\langle{#1}\bigr\rangle} 
\newcommand{\abs}[1]{\left\lvert{#1}\right\rvert}

\newcommand{\dd}{\mathrm{d}}
\newcommand{\ee}{\mathrm{e}}
\newcommand{\ii}{\mathrm{i}}

\newcommand{\kms}{\ensuremath{\mathrm{km\,s}^{-1}}}

\newcommand{\Mpc}{\ensuremath{\mathrm{Mpc}}}

\newcommand{\Msolar}{\ensuremath{\mathrm{M}_\odot}}

\newcommand{\arcmint}{\ensuremath{\mathrm{arcmin}}}

\newcommand{\Omegab}{\ensuremath{\Omega_\mathrm{b}}}
\newcommand{\Omegam}{\ensuremath{\Omega_\mathrm{m}}}

\newcommand{\vtheta}{\vect{\theta}}
\newcommand{\vvartheta}{\vect{\vartheta}}

\newcommand{\Map}{M_{\mathrm{ap}}}
\newcommand{\Nap}{\mathcal{N}}

\newcommand{\ngal}{n_{\mathrm{g}}}

\newcommand{\deltagal}{\delta_{\mathrm{g}}}
\newcommand{\kappagal}{\kappa_{\mathrm{g}}}

\newcommand{\meanngal}{\bar{n}_{\mathrm{g}}}

\newcommand{\deltam}{\delta_{\mathrm{m}}}

\begin{document}
\title{Confronting semi-analytic galaxy models with galaxy-matter correlations observed by CFHTLenS}

\author{
Hananeh Saghiha\inst{1}\thanks{Member of the International Max Planck Research School (IMPRS) for Astronomy and Astrophysics at the Universities of Bonn and Cologne.}
\and
Patrick Simon\inst{1}
\and
Peter Schneider\inst{1}
\and
Stefan Hilbert\inst{2,3}
}
\titlerunning{Confronting SAMs with observed galaxy-matter correlations}
\authorrunning{Saghiha et al.}

\institute{
Argelander-Institut f{\"u}r Astronomie, Universit{\"a}t Bonn, Auf dem H{\"u}gel
71, 53121 Bonn, Germany
\and
Exzellenzcluster Universe, Boltzmannstr. 2, 85748 Garching, Germany
\and
Ludwig-Maximilians-Universit{\"a}t, Universit{\"a}ts-Sternwarte, Scheinerstr. 1, 81679 M{\"u}nchen, Germany
}

\date{Received / Accepted }

\abstract{ 
Testing predictions of semi-analytic models of galaxy evolution against observations help to understand the complex processes that shape galaxies.
We compare predictions from the Garching and Durham models implemented on the Millennium Run with observations of galaxy-galaxy lensing (GGL) and galaxy-galaxy-galaxy lensing (G3L) for various galaxy samples with stellar masses in the range $0.5 \leq M_{*} / 10^{10}\, \Msolar  < 32$ and photometric redshift range $0.2 \leq z < 0.6$ in the Canada-France-Hawaii Telescope Lensing Survey (CFHTLenS).
We find that the predicted GGL and G3L signals are in qualitative agreement with CFHTLenS data. Quantitatively, the models succeed in reproducing the observed signals in the highest stellar mass bin ($16 \leq   M_{*} / 10^{10}\, \Msolar  < 32$) but show different degrees of tension for the other stellar mass samples. \TCB{The Durham models are strongly excluded at the 95\% confidence level by the observations as they largely over-predict the amplitudes of the GGL and G3L signals, probably because they predict too many satellite galaxies in massive halos.}
}
 
\keywords{gravitational lensing: weak -- large-scale structure of the Universe -- cosmology: observations -- galaxies: formation -- galaxies: evolution -- methods: numerical}

\maketitle

\section{Introduction}
\label{sect:introduction}

In the framework of the $\Lambda$CDM cosmology, galaxies and stars form from the gravitational collapse of baryonic matter inside dark matter halos. Semi-analytic models (SAMs) of galaxies are used to describe the connection between the resulting galaxy properties and the underlying distribution of dark matter \citep{1991ApJ...379...52W,1999MNRAS.303..188K,2001MNRAS.328..726S,2006RPPh...69.3101B}. Herein SAMs apply analytic prescriptions to approximate the complex processes of gas cooling, star formation, and feedback due to supernovae and active galactic nuclei. These prescriptions are calibrated to observations of galaxy properties such as the galaxy luminosity function or the Tully-Fischer relation using efficiency parameters and halo merger trees extracted from $N$-body simulations of structure formation \citep[e.g.,][]{2005Natur.435..629S,2012MNRAS.426.2046A}. \TCB{In comparison to hydrodynamical simulations \citep[e.g. ][]{2014MNRAS.444.1518V,2015MNRAS.446..521S} which are computationally expensive, SAMs enable a fast computation of model predictions for different parameters that describe galaxy physics. }

Gravitational lensing allows us to study the distribution of galaxies in relation to the matter density \citep[e.g. ][]{2001PhR...340..291B,schneider2006gravitational}. In the weak lensing regime, the tangential distortion of the image of a distant source galaxy or its ``shear'' may be measured as function of separation to foreground lenses to probe their correlation to the matter density field. This tangential shear is averaged over many lens-source pairs to obtain a detectable lensing signal. This galaxy-galaxy lensing (hereafter GGL) signal has been first detected by \citet{1996ApJ...466..623B}. The field of GGL has been growing rapidly since then thanks to larger surveys and more accurate shear measurements \citep[see e.g. ][]{2006MNRAS.368..715M,2011A&A...534A..14V,2012ApJ...744..159L,2014MNRAS.437.2111V,2015MNRAS.452.3529V,2016MNRAS.tmp..601V,2017MNRAS.465.4204C}. In essence, GGL measures the average projected matter density around lens galaxies. It thereby probes the statistical properties of dark matter halos in which galaxies reside. On small scales, GGL is dominated by the contribution from the host halo, but on larger scales, the neighboring halos also contribute to the signal.
 
\citet{2005A&A...432..783S} considered third-order correlations between lens galaxies and shear, called galaxy-galaxy-galaxy lensing (G3L). They defined two classes of three-point correlations: galaxy-shear-shear correlation function measured using triples composed of two sources and one lens galaxy, and the galaxy-galaxy-shear correlation function measured using triples comprising two lenses and one source galaxy. For this study, we consider only the lens-lens-shear correlations ($\cal{G}$), which measures the average tangential shear about lens pairs. The first detection of G3L was reported by \citet{2008A&A...479..655S} using the Red sequence Cluster Survey \citep[RCS,][]{2005ApJS..157....1G} data. The lens-lens-shear G3L essentially probes the stacked matter density around lens pairs in excess to the stack of two single lenses (Simon et al. 2012). Recently, the G3L signal was analyzed in the CFHTLenS by \citet{2013MNRAS.430.2476S}, where it was found that the amplitude of G3L increases with stellar mass and luminosity of the lens galaxies.

Measurements of GGL and G3L provide valuable data to test the ability of SAMs to correctly describe the connection between dark matter and galaxies as a function of scale and galaxy properties. The predictions for the expected lensing signals from SAMs needed for this comparison can be obtained by combining the simulated galaxy catalogs from the SAMs with outputs from gravitational lensing simulations using ray-tracing through the matter distribution of the underlying $N$-body simulation \citep[e.g. ][]{2009A&A...499...31H}. In \citet{2012A&A...547A..77S}, the G3L signal was computed for various galaxy models based on the Millennium Simulation \citep{2005Natur.435..629S}. There, the second- and third-order galaxy-matter correlation functions were represented in terms of aperture measures in the simulation, thereby allowing a straightforward comparison of different SAMs. \TCB{According to this study, G3L is a sensitive test for galaxy models and, in particular, different implementations of SAMs.}

In this paper, we compare SAM predictions of GGL and G3L to CFHTLenS data. \TCB{We consider four SAMs based on the Millennium Run: the Durham models by \citet[][hereafter B06]{2006MNRAS.370..645B} and \citet[][hereafter L12]{2012MNRAS.426.2142L}, and the Garching models by \citet[][hereafter G11]{2011MNRAS.413..101G} and \citet[][hereafter H15]{2015MNRAS.451.2663H}.}

The paper is organized as follows: Section~\ref{sect:theory} summarizes the formulation of GGL and G3L in terms of tangential shear and aperture statistics. In Sect.~\ref{sect:data}, we describe the complete data set and the method that we apply to select model galaxies from the SAMs. In Sect.~\ref{sect:results}, we compare the model predictions with lensing observations for various sub-samples of galaxies, based on redshift and stellar mass. We discuss our results in Sect.~\ref{sect:discussion}.

\section{Theory}
\label{sect:theory}
The lensing convergence for sources along the direction $\vvartheta$ is given by \citep[e.g.][]{schneider2006gravitational}
\begin{align}
\label{eq:kappagfunction}
\kappa (\vvartheta) &=\int \dd \chi_{\rm L} \, g(\chi_{\rm L}) \deltam(f_K(\chi_{\rm L})\vvartheta, \chi_{\rm L})
\quad\text{with}
\\
g(\chi_{\rm L}) &=
\frac{3\Omega_{\rm m}}{2D_{\rm h}^2} \frac{ f_K(\chi_{\rm L})}{a(\chi_{\rm L})}
\int_{\chi_{\rm L}}^{\infty} \dd \chi_{\rm S} \, p_{\rm S}(\chi_{\rm S}) \frac{f_K(\chi_{\rm S}-\chi_{\rm L})}{f_K(\chi_{\rm S})}\,.
\label{eq:gweight}
\end{align}
Here $\deltam(f_K(\chi)\vvartheta, \chi)$ is the matter density contrast of sources at comoving distance $\chi$ and comoving transverse position $f_K(\chi)\vvartheta$; $f_K(\chi)$ is the comoving angular diameter distance; $\Omega_{\rm m}$ is the cosmic mean matter density parameter; $D_{\rm
h}:=c/H_0$ is the Hubble length defined in terms of the vacuum speed of
light $c$ and the Hubble constant $H_0$; $a(\chi)$ is the scale
factor at radial comoving distance $\chi$. The sources have a radial distribution that is expressed by the
probability density function (PDF) $p_{\rm S}(\chi) $. The integral in Eq.~(\ref{eq:gweight}) is the lensing efficiency weighted by the source distribution.
The observed number density contrast
\begin{equation}
\label{eq:1}
\kappagal(\vvartheta)
= \frac{\ngal(\vvartheta) - \meanngal}{\meanngal}\,
\end{equation}
of lens galaxies on the sky with number density $n_{\rm g}(\vvartheta)$ and mean number density $\meanngal$ is a projection
of the 3D galaxy number density contrast $\deltagal(f_K(\chi)\vvartheta , \chi)$ weighted by the radial distribution $p_{\rm L} (\chi) \dd \chi $ of lenses:
\begin{equation}
\kappagal(\vvartheta) = \int \dd \chi \, p_{\rm L} (\chi) \deltagal(f_K(\chi)\vvartheta , \chi) .
\end{equation}

GGL probes the correlation of the inhomogeneities in the matter density and galaxy number density fields by cross-correlating the tangential shear in the source image and the position of the lens galaxy \citep[e.g.][]{2002ApJ...577..604H},
\begin{equation}
\bev{\gamma_{\rm t}}(\theta) = \ev{\kappagal (\vvartheta_{\rm L}) \gamma_{\rm t}(\vvartheta_{\rm S} ; \vvartheta_{\rm L})} ,
\end{equation}
where $\gamma_{\rm t}(\vvartheta_{\rm S} ; \vvartheta_{\rm L}) =  \rm{Re}[- \ee^{-2\ii \varphi} \gamma (\vvartheta_{\rm S})]$ is the tangential component of the shear, $\varphi$ is the polar angle of $\vtheta:=\vvartheta_{\rm L}-\vvartheta_{\rm S}$, $\theta = \abs\vtheta$, $\vvartheta_{\rm L}$ and $\vvartheta_{\rm S}$ are the lens and source galaxy positions.

The aperture mass
\begin{equation}
\Map(\vvartheta;\theta) = \int \dd^{2}\vartheta' \, U_\theta(\vert \vvartheta -
\vvartheta' \vert)\, \kappa(\vvartheta')
\label{eq:map}
\end{equation} 
employs convolution to obtain a smoothed convergence field for a circular aperture centered on $\boldsymbol{\vartheta}$ with angular scale $\theta$ \citep{1996MNRAS.283..837S}. The size of smoothing filter $U_\theta(\lvert \vvartheta \rvert)$  is given by the scale $\theta$. For a compensated filter function $U_\theta(\lvert \vvartheta \rvert)$ with 
\begin{equation}
\int\dd \vartheta\, \vartheta\, U_{\theta}(\vartheta) = 0, 
\end{equation}
the aperture mass can be written as
\begin{equation}
\Map(\vvartheta;\theta) = \int \dd^{2}\vartheta' \, Q_\theta(\vert \vvartheta -
\vvartheta' \vert)\, \gamma_{\rm t}(\vvartheta'),
\label{eq:mapshear}
\end{equation} 
with
\begin{equation}
Q_\theta(\vartheta) = \frac{2}{\vartheta^2} \int_{0}^{\vartheta} \dd \vartheta' \vartheta'  U_{\theta} (\vartheta') \, -\, U_{\theta} (\vartheta).
\end{equation}

In analogy to the aperture mass, the aperture number count of lenses is defined as
\begin{equation}
\Nap(\vvartheta; \theta)= \int \dd^{2} \vartheta' \, U_{\theta} (\vert \vvartheta
- \vvartheta' \vert) \, \kappa_{\mathrm{g}} ( \vvartheta')\,.
\label{eq:N}
\end{equation}
We use the exponential filter function introduced by \citet{1998A&A...334....1V},
\begin{equation}
U_{\theta}(\vartheta) = \frac{1}{\theta^{2}}\,u \left (\frac{\vartheta}{\theta} \right )\, ,
\end{equation}
with
\begin{equation*}
u(x) = \frac{1}{2\pi} \left ( 1 - \frac{x^{2}}{2} \right ) \exp \left
(\frac{-x^{2}}{2} \right) \, ,
\end{equation*}
to make predictions for the third-order moments of the aperture mass and aperture number count at zero lag and with equal aperture sizes:
\begin{equation}
\begin{split}
\label{eq:mmn}
&\ev{\Nap^2 \Map} (\theta) 
\equiv
  \ev{\Nap(\vvartheta;\theta) \Nap(\vvartheta;\theta) \Map(\vvartheta;\theta )}
\\&=
  \int \dd^{2} \vartheta_{1} \, U_{\theta} ( |\vvartheta_{1}|) \, \int \dd^{2}
\vartheta_{2} \, U_{\theta} ( |\vvartheta_{2}|) \, \int \dd^{2} \vartheta_{3} \,
U_{\theta} ( |\vvartheta_{3}|) \\ 
& \times
   \left\langle \kappa_{\mathrm{g}} ( \boldsymbol {\vartheta}_{1} )
\kappa_{\mathrm{g}} ( \boldsymbol {\vartheta}_{2} ) \kappa ( \boldsymbol
{\vartheta}_{3} ) \right\rangle \, .
\end{split}
\end{equation}

For homogeneous random fields such as $\kappa$ and $\kappa_{\rm g}$, the ensemble average $\ev{\Nap^2 \Map}$ is independent of the position of the centre of the aperture and can be calculated by (angularly) averaging the product $\Nap^2(\vvartheta;\theta) \Map(\vvartheta;\theta)$.
In our analysis we follow \cite{2012A&A...547A..77S} and calculate predictions for $\ev{\Nap^2 \Map}$ by averaging the product $\Nap^2 \Map$ over the simulated area.
Third-order aperture statistics can also be calculated from measurements of the lens-lens-shear correlator  $\cal G$ via integral transformations \citep{2005A&A...432..783S}. This has been done to obtain the CFHTLenS measurements.

\section{Data}
\label{sect:data}

\subsection{CFHTLenS galaxies}
\label{sect:CFHTLenSGalaxies}
CFHTLens is a multi-colour lensing survey \citep[][]{2012MNRAS.427..146H, 2013MNRAS.433.2545E,2013MNRAS.429.2858M}, incorporating $u^{*}g'r'i'z'$ multi-band data from the CFHT Legacy Survey Wide Programme. It covers $154\,\mathrm{deg}^2$ of the sky. Accurate photometry provided photometric redshifts of $7 \times10^{6}$ galaxies \citep{2012MNRAS.421.2355H}. The stellar masses of galaxies are estimated by fitting a model of the spectral energy distribution (SED) to the galaxy photometry. In this method, a set of synthetic SEDs are generated using a stellar population synthesis (SPS) model, and the maximum likelihood SED template that fits the observed photometry of a galaxy is obtained. Thus the SED fitting method relies on assumptions of the SPS models, star formation histories, initial mass function (IMF), and dust extinction models. The stellar masses of CFHTLenS galaxies are estimated using the SPS model of \citet{2003MNRAS.344.1000B} and assuming an IMF by \citet{2003PASP..115..763C}. By taking into account the error on the photometric redshift estimates as well as the uncertainties in the SED fitting, \citet{2014MNRAS.437.2111V} estimated that the statistical uncertainties on the stellar mass estimates of CFHTLenS galaxies are about 0.3 dex. 

In \citet{2013MNRAS.430.2476S}, the G3L analysis of CFHTLenS data is presented in terms of aperture statistics for a sample of source galaxies with $i' < 24.7$ and mean redshift of $z=0.93$, and lens galaxies brighter than $i' < 22.5$. The foreground sample is further subdivided in six stellar mass bins as given in Table~\ref{tab:Sm}. These stellar mass bins are then further split into two photometric redshift samples, $0.2 \leq z_{\textrm{ph}} < 0.44$ (``low-$z$'') and $0.44 \leq z_{\textrm{ph}} < 0.6$ (``high-$z$''). The redshift distribution of galaxies in these samples can be found in Fig.~5 of \citet{2013MNRAS.430.2476S}. We utilize these for the predictions of the lensing statistics.

{\renewcommand{\arraystretch}{1.5}
\begin{table}[htbp]
\caption{Binning in stellar mass of CFHTLenS galaxies for the low-$z$ and high-$z$ samples.}
\begin{tabular}{|c|c|c|}
\hline
stellar mass bin                &     selection    \\ \hline
sm1 &    $0.5 \leq   M_{*} / 10^{10} \Msolar  < 1$       \\
sm2 &        $ 1 \leq   M_{*} / 10^{10} \Msolar  < 2 $         \\
sm3 &   $ 2 \leq   M_{*} / 10^{10} \Msolar  < 4 $\\
sm4 &       $   4 \leq   M_{*} / 10^{10} \Msolar  < 8  $          \\
sm5 &    $8 \leq   M_{*} / 10^{10} \Msolar  < 16$ \\
sm6 &      $    16 \leq   M_{*} / 10^{10} \Msolar  < 32   $          \\ \cline{1-2}
\end{tabular}
\label{tab:Sm}
\end{table}
}
\subsection{Mock galaxies}
\label{sect:Mockgalaxies}
We use simulated lensing data obtained by a ray-tracing algorithm applied to the Millennium Simulation which is an $N$-body simulation that traces the evolution of $2160^{3}$ particles in a cubic region of comoving side length $500h^{-1}\Mpc$ from redshift $z=127$ to the present time \citep[MS,][]{2005Natur.435..629S}. The MS assumes a $\Lambda$CDM cosmology with parameters based on 2dFGRS \citep{2001MNRAS.328.1039C} and first-year WMAP data \citep{2003ApJS..148..175S}. These parameters are summarized in Table~\ref{tab:DMsimulations}.

{\renewcommand{\arraystretch}{1.5}
\begin{table}[htbp]
\centering
\caption{\TCB{Cosmological parameters for the assumed cosmology in the MS.}}
\begin{tabular}{|c|c|}
\hline
Parameters             & MS              \\
\cline{1-2}
$\Omega_\Lambda $      &  0.75           \\
$\Omegab $             &  0.045          \\
$\Omegam$              &  0.25            \\
$f_{\mathrm{b}}$       &  0.17             \\
$\sigma_{8} $          &  0.9              \\
$n_{\mathrm{s}}$       &  1.0               \\
$H_{0}[\kms\Mpc^{-1}]$ & 73                 \\
\cline{1-2}
\end{tabular}
\label{tab:DMsimulations}
\end{table}
}

\TCB{We use galaxy catalogs from B06, G11, L12, and H15 implemented on the MS.}\footnote{One prominent improvement in H15 is that the simulations are rescaled to the Planck cosmology according to the method described in \citet{2010MNRAS.405..143A} and \citet{2015MNRAS.448..364A}. However, here we use the H15 model adjusted to the original MS cosmology.} All these \TCB{four} models use similar treatments for basic physical baryonic processes such as gas cooling, star formation and feedback from supernovae and AGNs, but they differ in various details. We refer to some of these differences later in the paper.

The gravitational lensing in the Millennium simulation is computed by the multiple-lens-plane ray-tracing algorithm of \citet[][]{2009A&A...499...31H} in 64 fields of view of $4 \times 4\,\mathrm{deg}^2$ each. 
The resulting synthetic data include the convergence and shear (on regular meshes of $4096^2$ pixels) of sources at a set of redshifts given by the output times of the simulation snapshots. These are then combined into convergence and shear fields for the CFHTLenS redshift distribution. Furthermore, the data contains the image positions, redshifts, stellar masses, and various other galaxy properties of the galaxies computed by the SAMs.

\begin{figure}
\centerline{\includegraphics[width=90mm]{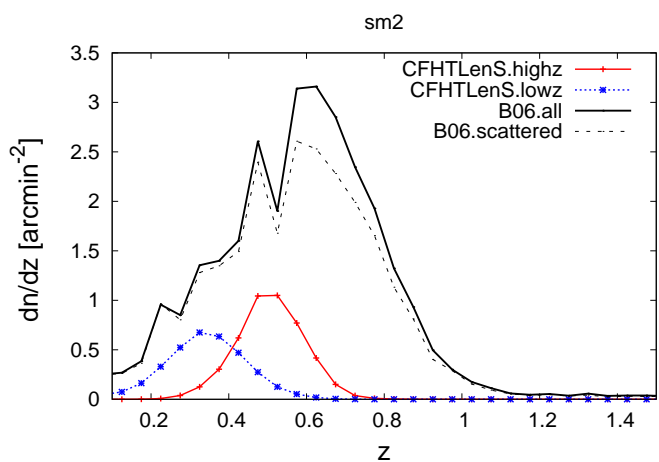}}
\caption{Number density distribution per unit solid angle and redshift interval of flux-limited galaxies in sm2 (the total area below each curve is the total number density of galaxies). The blue and red curves show the distribution of CFHTLenS galaxies in the low-$z$ and high-$z$ samples, respectively. The solid black curve represents all galaxies of the B06 model above the flux limit and with stellar mass in the sm2 bin. The dashed black curve shows the distribution when also applying a random error to the stellar masses of the B06 galaxies. }
\label{fig:nz_Bower_los801_sm2}
\end{figure}

We generate mock galaxy samples similar to the lens samples observed in CFHTLenS following three steps.

(i)
We convert the SAM magnitudes to the Megacam AB magnitudes in CFHTLenS. We convert the SDSS AB magnitudes of \TCB{G11/H15} to Megacam AB magnitude $i_{\text{AB}}^{'}$ by applying the conversion relation from \citet{2013MNRAS.433.2545E}:
\begin{equation}
	i_{\text{AB}}^{'} = i_{\text{AB}} - 0.085(r_{\text{AB}} - i_{\text{AB}}) \, .
\end{equation}
We convert the SDSS Vega magnitudes of \TCB{B06/L12} to CFHTLenS AB magnitudes using:\footnote{\texttt{http://www.cfht.hawaii.edu/Instruments/Imaging/\\MegaPrime/specsinformation.html}}
\begin{equation}
	i_{\text{AB}}^{'} = (i_{\text{Vega}} + 0.401) - 0.085\left ( \left [r_{\text{Vega}} + 0.171\right ] - \left [i_{\text{Vega}} + 0.401 \right ] \right )\,.
\end{equation}
We then select lens galaxies brighter than $i_{\text{AB}}^{'} < 22.5$. The redshift distribution of all flux-limited galaxies from the B06 model that fall into the stellar-mass bin sm2 is shown in Fig.~\ref{fig:nz_Bower_los801_sm2}.

(ii)
In order to emulate the CFHTLenS error of stellar masses, we randomly add Gaussian noise with RMS 0.3 dex to the stellar mass $\log M_{*}$ in the mocks. The resulting redshift distribution of galaxies in B06 is also shown in Fig.~\ref{fig:nz_Bower_los801_sm2}.

(iii) 
As can be seen in Fig.~\ref{fig:nz_Bower_los801_sm2}, the redshift distribution of model galaxies differs from that of CFHTLenS (the dashed black curve compared to the red or blue curves in Fig.~\ref{fig:nz_Bower_los801_sm2}). To select a realistic simulated sample, the mock samples must have the same redshift distributions as the corresponding CFHTLenS samples to produce the same lensing efficiency. Therefore, in the last step, we use a rejection method to reproduce the redshift distribution of galaxies in CFHTLenS. In this step, we randomly discard a galaxy at redshift $z$ from the mock sample if 
\begin{equation}
x > \frac{\dd n_{\textrm{SAM}}/ \dd z}{\dd n_{\textrm{CFHTLenS}}/ \dd z}
\end{equation}
is satisfied for a random number in the range 0 to 1. The distribution of selected galaxies in the low-$z$ and high-$z$ samples are not shown in Fig.~\ref{fig:nz_Bower_los801_sm2} since they are practically identical to the corresponding CFHTLenS distributions.

\begin{figure}
\centerline{\includegraphics[width=90mm]{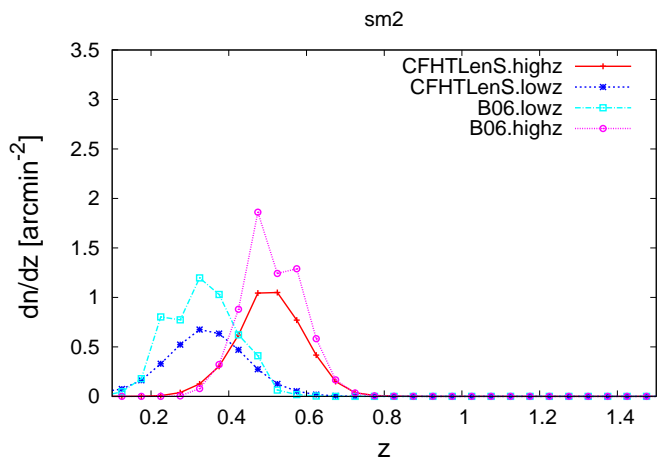}}
\caption{Similar to Fig.~\ref{fig:nz_Bower_los801_sm2}, the blue and red curves show the distribution of CFHTLenS galaxies in the low-$z$ and high-$z$ samples, respectively. The curves labeled as ``B06.lowz'' (cyan) and ``B06.highz'' (magenta) correspond to a sample of galaxies selected from the distribution shown by the dashed black curve in Fig.~\ref{fig:nz_Bower_los801_sm2} after adding the error of photometric redshifts in CFHTLenS to the mock redshifts.}
\label{fig:nz_Bower_los801_sm2_test}
\end{figure}

One should note that in the method described above, we have not included the error in the 
photo-$z$ estimation. However, including such an uncertainty has no effect on the statistical properties of the SAM galaxy distributions. 
Indeed, Fig.~\ref{fig:nz_Bower_los801_sm2_test} shows the true redshift distribution of B06 galaxies for low-$z$ and high-$z$ samples after including an emulated photo-$z$ error, and
after applying the same photo-$z$ cuts as in CFHTLens. For this, we choose galaxies from the ``B06.scattered'' distribution (dashed black curve in Fig.~\ref{fig:nz_Bower_los801_sm2}) and add a random Gaussian photo-$z$ error with RMS $0.04(1+z)$  \citep{2012MNRAS.421.2355H}. The PDF of the true redshifts of low-$z$ and high-$z$ samples are labelled ``B06.lowz'' and ``B06.highz'' in Fig.~\ref{fig:nz_Bower_los801_sm2_test}. Despite having slightly different amplitudes, these distributions have similar shapes as the ``CFHTLenS.lowz'' (blue) and ``CFHTLenS.highz'' (red) distributions, respectively. 
After applying the rejection method (step iii) on those distributions, one obtains mock galaxy samples 
with the same statistical properties of the mock galaxy samples that we produce following the three steps described previously.

\section{Results}
\label{sect:results}

\subsection{GGL}
\label{sect:GGL}

\begin{figure*}
 \centering
 \begin{subfigure}[b]{0.85\textwidth}
     \includegraphics[width=\textwidth]{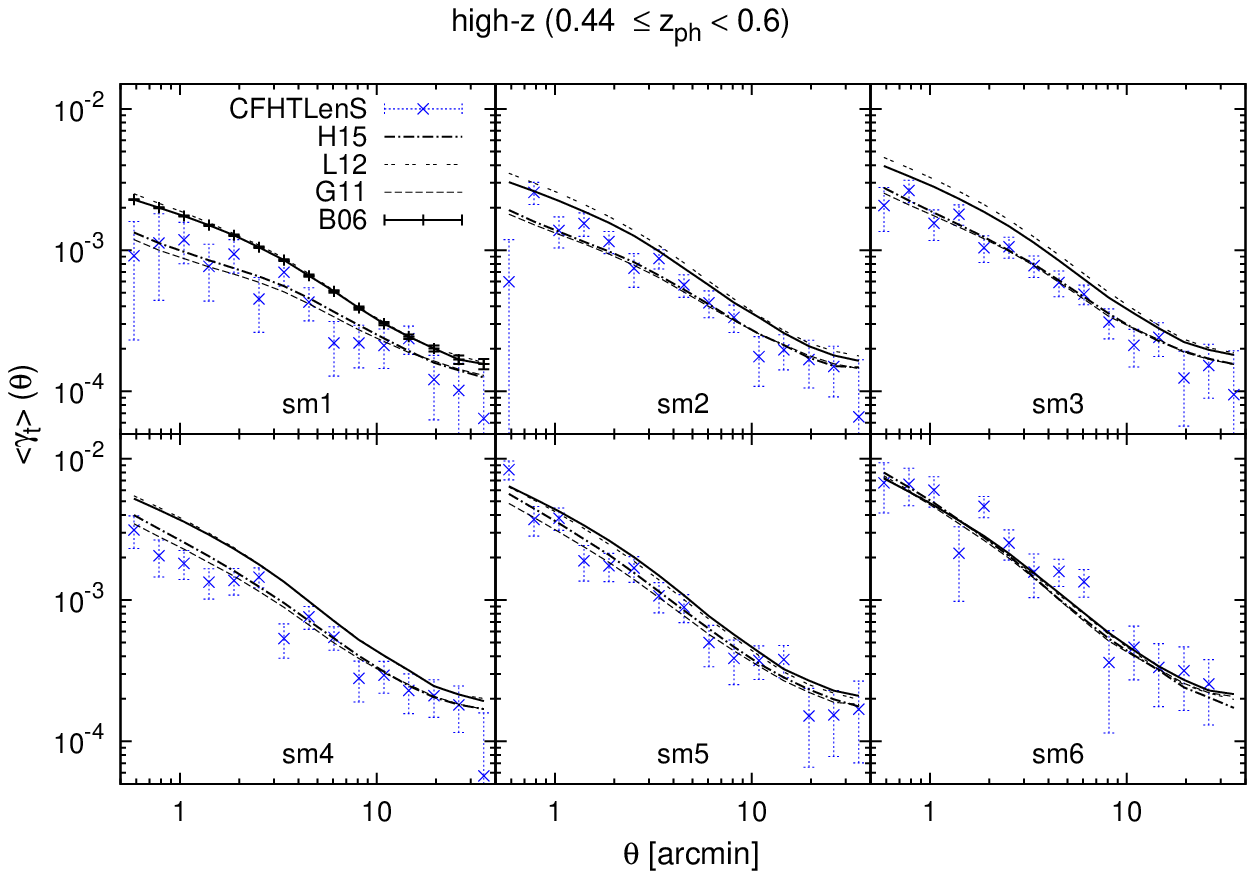}
     \label{fig:TangentialShear:highz}
 \end{subfigure}
 ~
 \begin{subfigure}[b]{0.85\textwidth}
     \includegraphics[width=\textwidth]{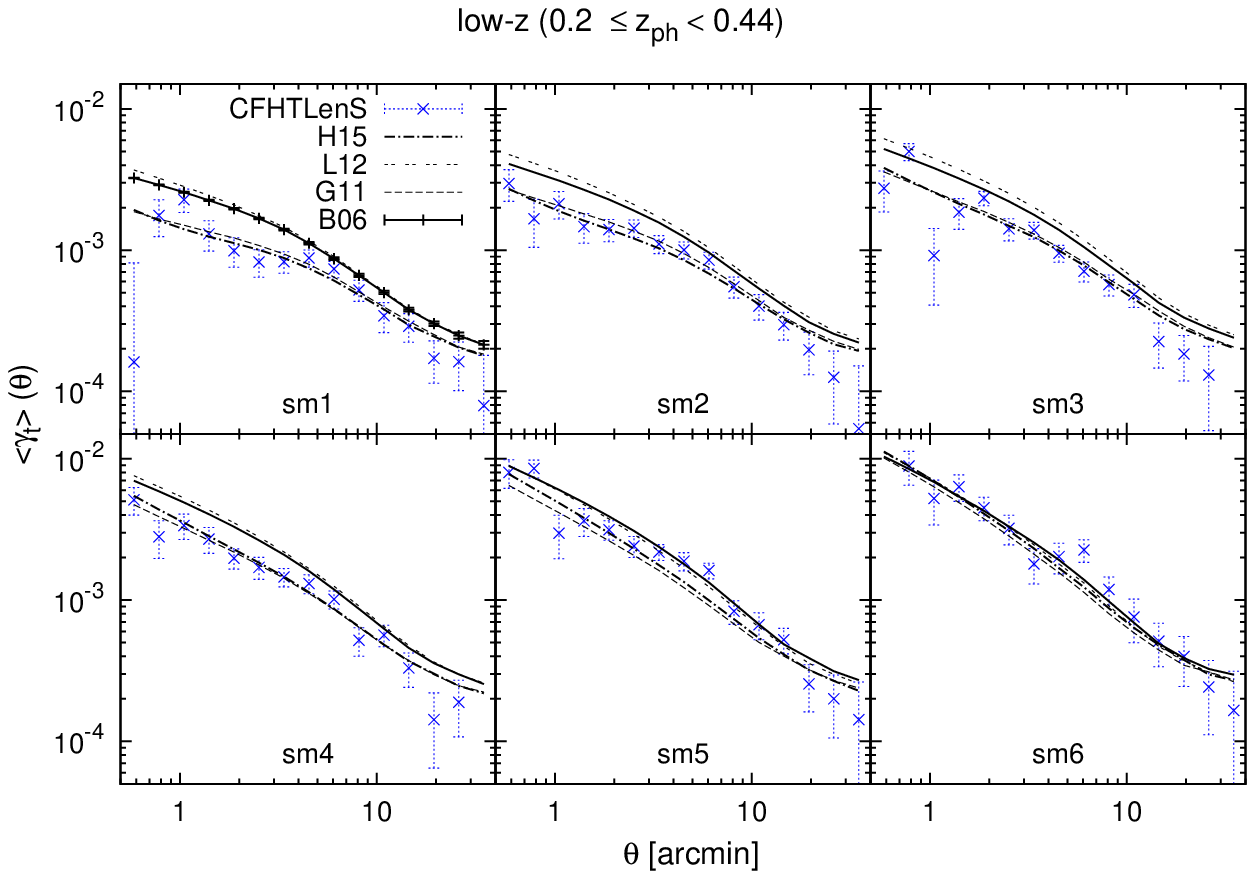}
     \label{fig:TangentialShear:lowz}
 \end{subfigure}
 \caption{
GGL as function of projected separation for the six stellar mass samples according to Table~\ref{tab:Sm}. The top panel corresponds to the high-$z$ sample and the bottom panel to low-$z$. The data points with error bars (indicating the standard error of the mean over 129 fields) show the CFHTLenS measurements, \TCB{which are compared to the predictions by B06 (solid lines), G11 (dashed lines), L12 (double dashed lines), and H15 (dash dotted lines).} The B06 predictions for sm1 show the error of the mean over 64 fields.
}
 \label{fig:TangentialShear}
\end{figure*}

Fig.~\ref{fig:TangentialShear} shows the azimuthally averaged tangential shear $\bev{\gamma_{\rm t}}(\theta)$ for an angular range of 0.5 to 35 \ensuremath{\mathrm{arcmin}} as measured in CFHTLenS in comparison to the SAM predictions. The samples are split in stellar mass and redshift.
For both CFHTLenS data and model galaxies, the amplitude of the GGL signal increases with stellar mass. For a given stellar mass bin, the amplitudes of the observed and simulated signals decrease when increasing the lens-source separation $\theta$.

To quantify the significance of the difference between model predictions and CFHTLenS measurements of  $\bev{\gamma_{\mathrm{t}}}$, we compute
\begin{equation}
 \chi^{2} =  \left (\mathbf{d}^{\mathrm{sam}} - \mathbf{d}^{\mathrm{obs}}  \right)^{\rm T}\mathbf{C}^{-1} \left (\mathbf{d}^{\mathrm{sam}} - \mathbf{d}^{\mathrm{obs}}  \right) \, ,
\end{equation}
where $\mathbf{d}^{\mathrm{sam}}$ and $ \mathbf{d}^{\mathrm{obs}} $ are data vectors containing the SAMs predictions and CFHTLenS measurements, respectively. The covariance matrix $\mathbf{C}$ of the difference signal is $\mathbf{C}^{\mathrm{sam}} + \mathbf{C}^{\mathrm{obs}}$ since SAMs and CFHTLenS measurements are uncorrelated. \TCB{Here $\mathbf{C} ^{\mathrm{sam}}$ is the field-to-field covariance of SAMs estimated using $64$ simulated fields, and $\mathbf{C} ^{\mathrm{obs}}$ is the Jackknife covariance of CFHTLenS measurements using $N^{\mathrm{obs}} = 129$ fields-of-view of $1 \times 1\,\mathrm{deg}^2$ each. We construct $N^{\mathrm{obs}}$ Jackknife samples and store the mean of the combined samples excluding the $i$th field in the data vector $\mathbf{d}_{i}$. The vector $\mathbf{d}$ shall be the average of all $\mathbf{d}_{i}$ vectors. The Jackknife covariance of the sample mean is then   
\begin{equation}
\mathbf{C}^{\mathrm{obs}} = \frac{N^{\mathrm{obs}} -1}{N^{\mathrm{obs}}} \sum_{i=1}^{N^{\mathrm{obs}}}  \left(\mathbf{d} - \mathbf{d}_{i}\right) \left(\mathbf{d} - \mathbf{d}_{i}\right)^{\rm T} \,.
\end{equation}
}

For our $\chi^{2}$ test, we have $\mathbf{C} \simeq \mathbf{C}^{\mathrm{obs}}$ because the elements of the SAMs covariance matrix are negligible in comparison with the elements of the CFHTLenS covariance matrix. We apply the estimator of \citet{2007A&A...464..399H} to obtain an estimator for the inverse of the covariance $\mathbf{C}^{-1}$ for $\mathbf{C}^\mathrm{obs}$,
\begin{equation}
\mathbf{C}^{-1} = \frac{N^{\mathrm{obs}} - N^{\mathrm{d}}  -2}{N^{\mathrm{obs}} -1} (\mathbf{C}^\mathrm{obs})^{-1},
\end{equation} 
\TCB{when $N^{\mathrm{d}}<N^{\mathrm{obs}}-2$.  $N^{\mathrm{d}}$ is the number of data points} and $N^{\mathrm{obs}} = 129$ is the number of Jackknife realizations used for $\mathbf{C}^\mathrm{obs}$. 

{\renewcommand{\arraystretch}{1.3}
\begin{table}[htbp]
\caption{$\chi^{2}$-test with \TCB{$N^{\mathrm{d}}=15$} degrees of freedom applied to measurements of GGL shown in Fig.~\ref{fig:TangentialShear}. Each number quotes the \TCB{reduced $\chi^{2}/N^{\mathrm{d}}$} for the corresponding model and stellar mass bin. Bold values indicate a tension between CFHTLenS and a SAM at 95\% confidence level.}
\centering
\resizebox{0.47\textwidth}{!}
{
\begin{tabular}{|ccccc|cccc|}
\hline
       & \multicolumn{4}{c|}{low-$z$} & \multicolumn{4}{c|}{high-$z$}    \\         
       &  G11     & H15    & B06  &\TCB{L12}  &  G11    & H15    & B06   &\TCB{L12}            \\ \cline{1-9}
sm1 &   1.25          &   1.27  &  \textbf{6.07}   &     \TCB{\textbf{7.80}}             &0.68    & 0.71   &     \textbf{2.61}   &      \TCB{\textbf{2.98}}                    \\
sm2 &   0.54         &  0.56    &  \textbf{4.58}   &       \TCB{\textbf{7.97}}             & 1.26    & 1.23      &   \textbf{3.00}   &   \TCB{\textbf{ 4.58 }}                       \\
sm3 &   \textbf{2.43}         &  \textbf{2.36}   &  \textbf{6.78}&  \TCB{\textbf{11.49}}    & 0.45    & 0.46   &   \textbf{3.75}   &    \TCB{\textbf{6.59 }}                         \\
sm4 &  0.93          &  1.11    & \textbf{5.37}                  & \TCB{\textbf{7.08}}     & 1.11    & 1.64   &   \textbf{7.09}   &    \TCB{\textbf{7.71 }}                      \\
sm5 &   \textbf{2.30}         &  \textbf{1.77}   & \textbf{2.12} &  \TCB{\textbf{1.84}}    & 1.06    & 0.88   &   1.62            &    \TCB{1.28}                 \\
sm6 &   1.26         &   0.97   & 1.00                 &  \TCB{0.92}     & \textbf{1.79}     & 1.63  &   1.49   &     \TCB{1.55}                          \\ \cline{1-9}
\end{tabular}}
\label{tab:chisquareGGL}
\end{table}
}

For $N^{\mathrm{d}}=15$ degrees of freedom, a tension between CFHTLenS and the SAM predictions with 95\% confidence is given by values of $\chi^2/15>1.67$ (written in bold in Table~\ref{tab:chisquareGGL}). The results from Table~\ref{tab:chisquareGGL} clearly show that \TCB{the B06/L12 models are in tension for all stellar mass bins except for sm5 in the high-$z$ sample and sm6. In comparison, the G11/H15 models are consistent with the observations apart from possible tensions for sm3 and sm5 at low-$z$.}

\TCB{To quantify the overall difference in GGL between the SAMs and CFHTLenS, we combine the measurements of all stellar mass samples and test for a vanishing difference signal consisting of $N^{\mathrm{d}}=90$ data points. Since data points between different stellar masses are correlated, we estimate a new $90\times90$ covariance by Jackknifing the combined bins in 129 CFHTLenS and for the 64 mock fields. The results of the $\chi^2$ test are presented in Table~\ref{tab:chisquareCombinedGGL}. A tension between model and observation is now indicated by $\chi^2/90>1.26$ (95\% confidence level). According to the $\chi^2$ test, only the predictions of H15 for the low-$z$ sample are in agreement with the CFHTLenS. The B06/L12 models show the strongest tension.}
{\renewcommand{\arraystretch}{1.3}
\begin{table}[htbp]
\caption{\TCB{Similar to Table~\ref{tab:chisquareGGL} but for all stellar mass samples combined. Therefore, the degrees of freedom are $N^{\mathrm{d}}=90$ . }}
\centering
\resizebox{0.47\textwidth}{!}
{
\begin{tabular}{|ccccc|cccc|}
\hline
  & \multicolumn{4}{c|}{\TCB{low-$z$}} & 
\multicolumn{4}{c|}{\TCB{high-$z$}}    \\
  & \TCB{G11}  & \TCB{H15}  & \TCB{B06}    &\TCB{L12}      &  \TCB{G11}    & \TCB{H15}  & \TCB{B06} & \TCB{L12}
\\ \cline{1-9}
\TCB{sm1-6} & \TCB{\textbf{1.37}}     &  \TCB{1.22}  & \TCB{\textbf{3.98}} & \TCB{\textbf{5.50}}&  \TCB{\textbf{1.29}}   & \TCB{\textbf{1.35}}   &  \TCB{\textbf{2.88}} & \TCB{\textbf{3.58}}      \\  \cline{1-9}
\end{tabular}}
\label{tab:chisquareCombinedGGL}
\end{table}
}

\subsection{G3L}
\label{sect:GGGL}
\begin{figure*}[htbp]
 \centering
 \begin{subfigure}[b]{0.85\textwidth}
     \includegraphics[width=\textwidth]{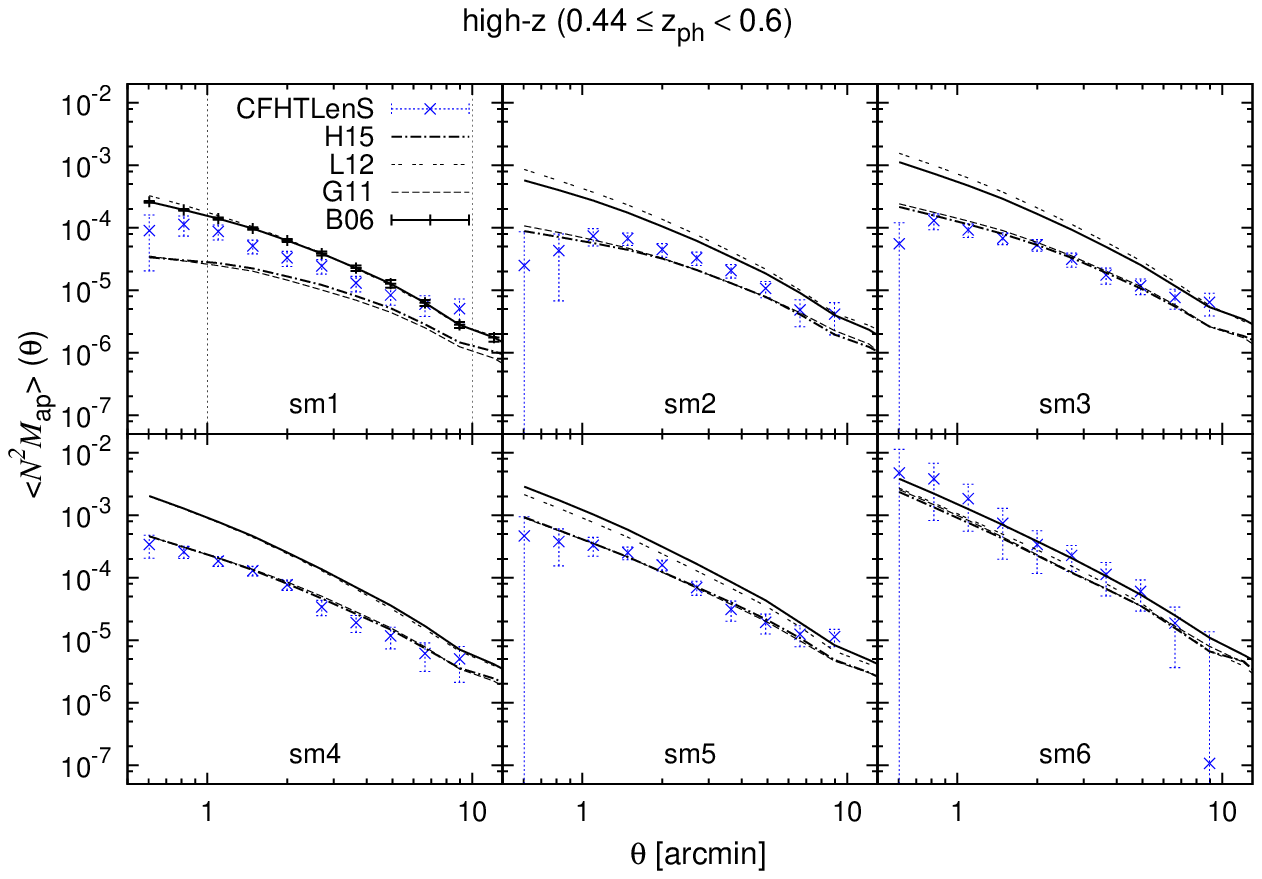}
     \label{fig:NNM:highz}
 \end{subfigure}
 ~
 \begin{subfigure}[b]{0.85\textwidth}
     \includegraphics[width=\textwidth]{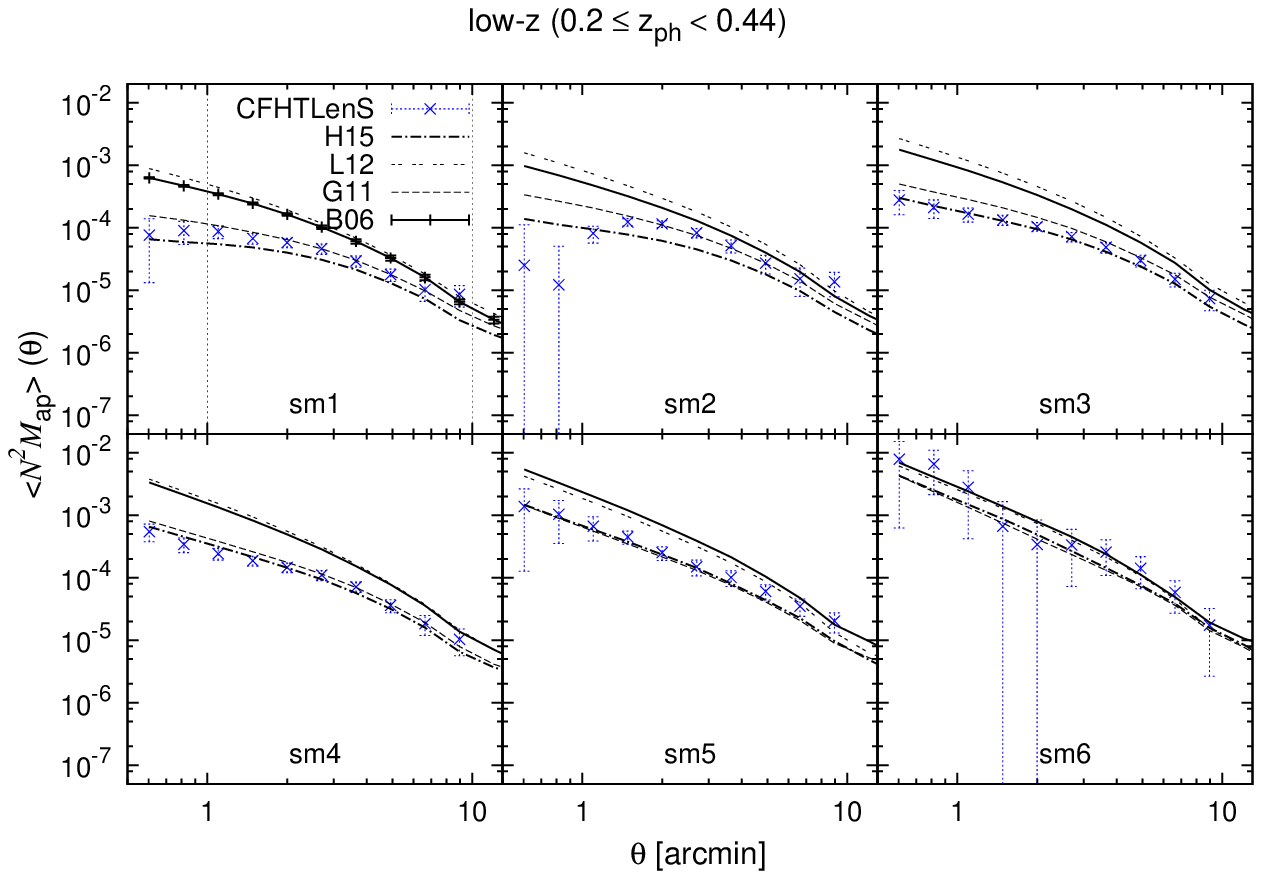}
     \label{fig:NNM:lowz}
 \end{subfigure}
 \caption{ Measurements of the G3L aperture statistics as function of aperture scale $\theta$ in CFHTLenS (blue symbols) and SAMs (black curves). Measurements are presented for various stellar mass and redshift (high-$z$ and low-$z$) samples. Error bars indicate the standard error of the mean. The dotted vertical lines show the limits of the range used for our $\chi^{2}$ analysis.
}
 \label{fig:NNM}
\end{figure*}

The $\bev{\Nap^2 \Map}(\theta)$ values measured in CFHTLenS for the low-$z$ and high-$z$ samples in all stellar mass bins are shown in Fig.~\ref{fig:NNM}. Also shown there are the predictions from the SAMs. The observed $\bev{\Nap^{2}\Map} (\theta)$ signal is dominated by the ``transformation bias" below 1 $\arcmint$ and above 10 $\arcmint$ \citep{2008A&A...479..655S}. This bias is caused by galaxy blending and the finite size of the observed field, thus leading to insufficient sampling of the three-point correlation function \citep{2006A&A...457...15K}. Therefore, only data points between $1'<\theta<10'$ are used for comparison, indicated by the dashed vertical lines in the top left panel. We retain \TCB{$N^{\mathrm{d}}=8$} data points for each stellar mass and redshift bin.

{\renewcommand{\arraystretch}{1.3}
\begin{table}[htbp]
\caption{\TCB{$\chi^{2} / N^{\mathrm{d}}$ for $N^{\mathrm{d}} = 8$ degrees of freedom applied to measurements of G3L shown in Fig.~\ref{fig:NNM}. Only data points between $1'<\theta<10'$ were used for this test. Bold values indicate a tension between CFHTLenS and a SAM at 95\% confidence level.}
}
\centering
\resizebox{0.47\textwidth}{!}
{
\begin{tabular}{|ccccc|cccc|}
\hline
  & \multicolumn{4}{c|}{low-$z$} & \multicolumn{4}{c|}{high-$z$}    \\
  & G11   & H15  & B06   & \TCB{L12}        &  G11    & H15  & B06  & \TCB{L12}  \\ \cline{1-9}
sm1 &   0.88 &   1.33  & \textbf{27.35}                   & \TCB{\textbf{48.27}} &  \textbf{2.07} &  1.73 &  \textbf{3.20} & \TCB{\textbf{3.77}} \\
sm2 &    \textbf{5.66}  &  \textbf{4.13} & \textbf{33.18} & \TCB{\textbf{86.71}} &1.27 & 1.39 & \textbf{11.19}& \TCB{\textbf{24.57}}\\
sm3 &   1.69    & 0.29 & \textbf{38.80} &         \TCB{\textbf{93.54}}            &     1.15 & 0.99 & \textbf{35.59}& \TCB{\textbf{69.95}}\\
sm4 &   1.43  &  0.81 & \textbf{68.81} &         \TCB{\textbf{94.11}}              &  1.31 &  1.19 &\textbf{44.99}& \TCB{\textbf{41.61}} \\
sm5 &    1.24  &  1.26 &  \textbf{8.67} &        \TCB{\textbf{5.23}}            &1.57 & 1.57 & \textbf{7.15}& \TCB{\textbf{3.36}}\\
sm6 &     0.59 &  0.63 &  0.71 &          \TCB{0.54}                    & 1.70 & 1.59 &  1.48 & \TCB{1.43}\\  \cline{1-9}  
\end{tabular}}
\label{tab:chisquare}
\end{table}
}

Our measurements show that the \TCB{G11/H15} predictions agree better with CFHTLenS than \TCB{B06/L12. The B06/L12 models over-predict} the $\bev{\Nap^{2}\Map} (\theta)$ signal in all but the highest stellar mass bin. In addition, the tension between \TCB{B06/L12} and CFHTLenS is more prominent for G3L than in the GGL measurements \TCB{as can be deduced from the $\chi^{2}$ values in Table~\ref{tab:chisquare}}. Model measurements with $\chi^{2}/8 > 1.94$, i.e. a SAM signal inconsistent with \TCB{CFHTLenS are written in bold (95\% confidence level). }

{\renewcommand{\arraystretch}{1.3}
\begin{table}[htbp]
\caption{\TCB{Similar to Table~\ref{tab:chisquare} but for all stellar mass samples combined, i.e., $N^{\mathrm{d}}=48$. } }
\centering
\resizebox{0.47\textwidth}{!}
{
\begin{tabular}{|ccccc|cccc|}
\hline
  & \multicolumn{4}{c|}{low-$z$} & 
\multicolumn{4}{c|}{high-$z$}    \\
  & G11  & H15  & B06    &\TCB{L12}      &  G11    & H15  & B06 & \TCB{L12}
\\ \cline{1-9}
sm1-6 & \textbf{2.56}     &  1.34  & \textbf{27.73} &\TCB{\textbf{51.77}} &  \textbf{2.18}   & \textbf{2.08}   &  \textbf{12.88} & \TCB{\textbf{16.42}}      \\  \cline{1-9}
\end{tabular}}
\label{tab:chisquareCombined}
\end{table}
}

\TCB{Similar to GGL, we combine the G3L measurements of
all stellar mass samples to quantify the overall difference between SAMs
and CFHTLens ($N^{\mathrm{d}}=48$). We estimate a new $48\times48$ covariance by Jackknifing the combined bins in 129 CFHTLenS and for the 64 mock fields.} The results of the $\chi^2$ test are presented in Table~\ref{tab:chisquareCombined}. A tension between model and observation is now indicated by $\chi^2/48>1.35$ (95\% confidence level). According to the $\chi^2$ test, only the predictions of H15 for the low-$z$ sample are in agreement with the CFHTLenS.  

\subsection{Stellar mass distribution}
\label{sect:SMF}
\begin{figure}[htbp]
  \begin{subfigure}{0.53\textwidth}
      \includegraphics[width=\textwidth]{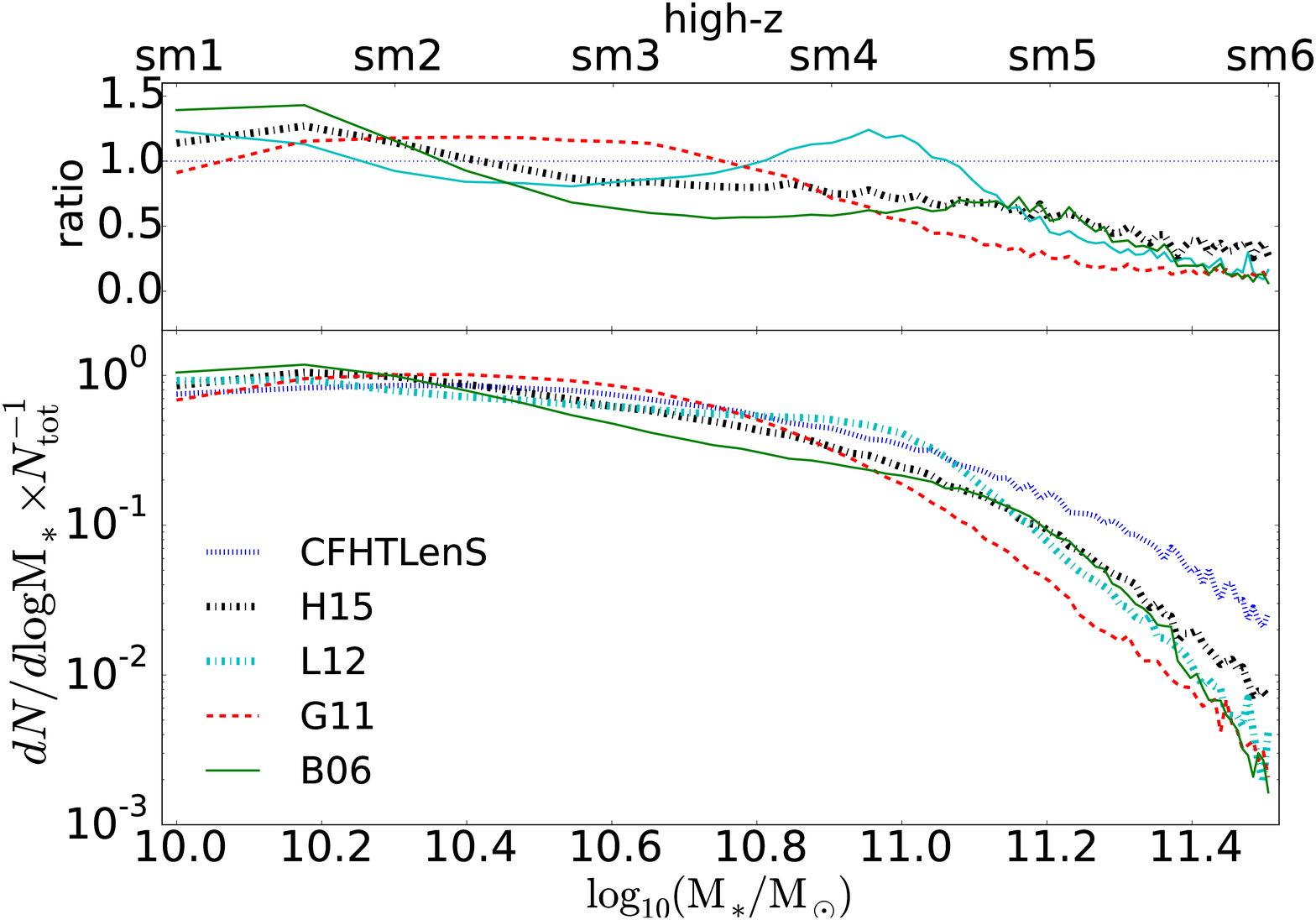}
  \end{subfigure}
 \\
  \begin{subfigure}{0.53\textwidth}
     \includegraphics[width=\textwidth]{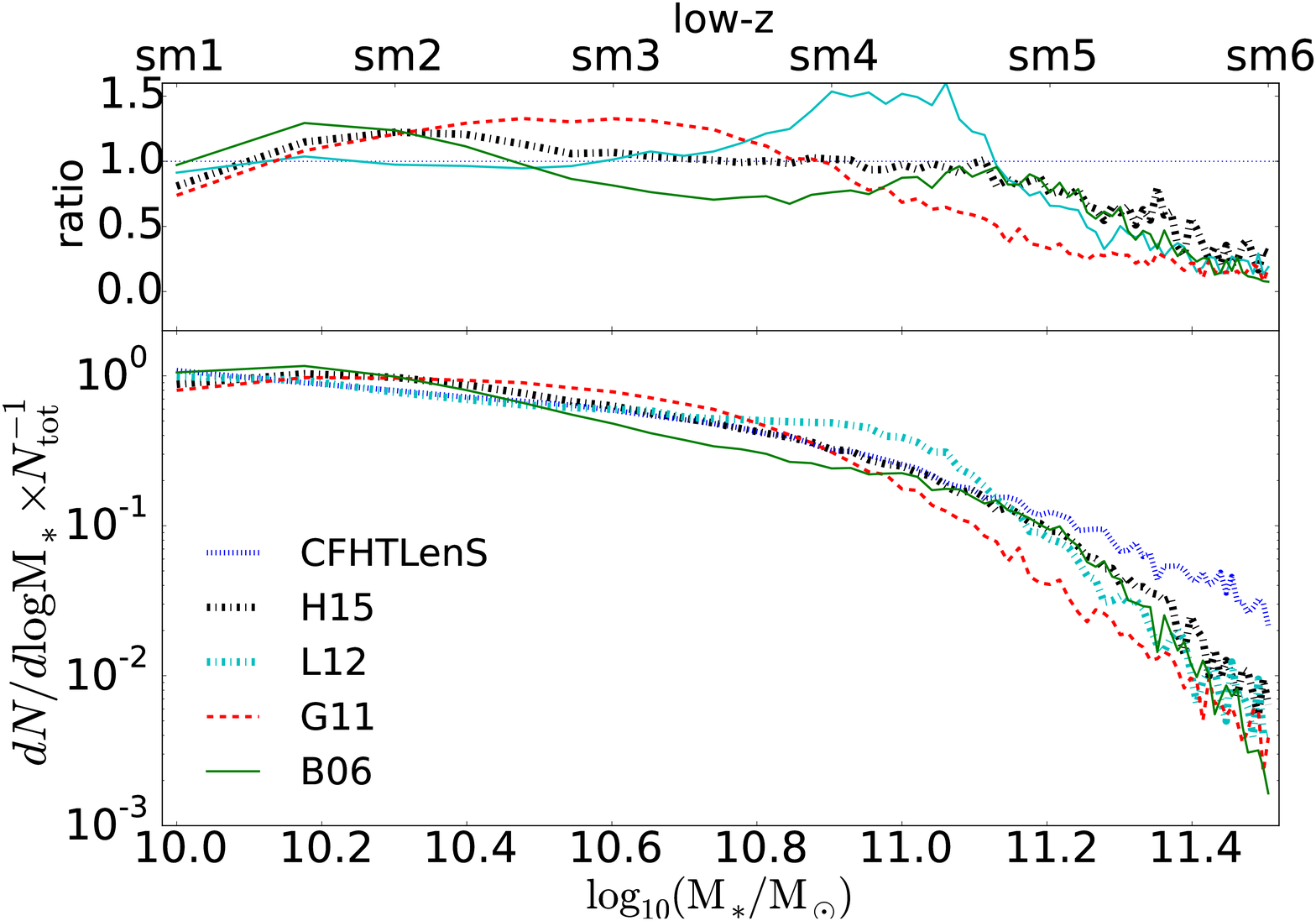}
  \end{subfigure}
  \caption{The stellar mass function of galaxies normalized with the total number of galaxies in all three SAMs and CFHTLenS. We used a sample of sm1 to sm6 combined and repeated the three-step selection in Sect.~\ref{sect:Mockgalaxies} to produce high-$z$ (top figure) and low-$z$ (bottom figure) subsamples. The top of each panel shows the ratio between the SAM and CFHTLenS stellar mass function.
	}
\label{fig:SMFratio}
\end{figure}

Given the foregoing results, we test if the stellar masses of galaxies are systematically different between the SAMs.
For this purpose, we selected mock galaxies as described in Sect.~\ref{sect:Mockgalaxies} for a broad stellar mass bin including $ M_{*}$ from $5\times10^{10}$ to $3.2\times10^{11}\rm M_\odot$ and plot the resulting distribution $\dd N / \dd \log  M_{*} $ in Fig.~\ref{fig:SMFratio} for 63 bins in $ {M}_{*}$. The number of galaxies $N$ in a bin is normalized by the total number $N_{\text{tot}}$ in the plotted range. The ratios of the model predictions and the CFHTLenS results are shown in the upper panel of each plot. For comparison, we indicate on top of the plot labels corresponding  to stellar mass samples sm1-sm6. The SAMs results are lower than that of CFHTLenS in high stellar mass bins, and their distributions drop more quickly. In low stellar mass bins, there are differences between G11 and B06 compared to H15, for instance there is a dip for B06 in the range sm2 to sm4 compared to H15. The stellar mass distribution of H15 is the closest to CFHTLenS.

\newpage
\section{Discussion}
\label{sect:discussion}
In this work, we have studied second- and third-order galaxy-mass correlation functions in terms of average tangential shear $\bev{\gamma_{\mathrm{t}}}$ and aperture statistics $\bev{\Nap^{2}\Map}$, respectively. We used mock galaxies from \TCB{the Durham models (B06 and L12), and the Garching models (G11 and H15)} which are SAMs implemented on the Millennium Simulation. We compared our results with the observational results of CFHTLenS for galaxies binned in stellar mass within $0.6 <   M_{*} / 10^{10} \Msolar  < 32$ and redshift within $0.2 \leq z_{\rm {ph}} \leq 0.6$. In addition, all lens galaxies are subject to a flux limit of $i_{\text{AB}}^{'} < 22.5$.

Our results indicate that not all models can reproduce the GGL and G3L observations although there is an overall qualitative agreement between the models and CFHTLenS as visible in the Fig. \ref{fig:TangentialShear} and Fig. \ref{fig:NNM}. All models best agree among each other and with CFHTLenS for sm6, i.e., for stellar masses of $\sim2\times10^{11}\,\rm M_\odot$. However, the uncertainties of the CFHTLenS results are also largest here. At lower stellar masses, see Tables~\ref{tab:chisquareGGL} and \ref{tab:chisquare}, \TCB{the Durham models clearly over-predict the amplitude of both GGL and G3L so that these models} can be decisively excluded at the 95\% confidence level. The agreement between \TCB{the Garching models} and CFHTLenS, on the other hand, is good although the overall comparison to G3L still indicates some tension in Table~\ref{tab:chisquareCombined}. The fit of the more recent H15 is slightly better compared to G11.
We also find from our $\chi^2$ values that G3L has more discriminating
power than GGL on the same data, as anticipated in \citet{2012A&A...547A..77S}. \TCB{This may be understood by G3L being more sensitive to the clustering amplitude of the lenses when compared to GGL. For a linear deterministic bias $b$ of the lens number density, the G3L signal is proportional to $b^2$ whereas the signal is proportional to $b$ for GGL. }

A systematically high galaxy-matter correlation in \TCB{the Durham models} might indicate that the stellar masses of galaxies \TCB{in these models} are systematically higher compared to \TCB{the Garching models}. Such bias could impact the matter environment, clustering, and hence GGL and G3L of stellar-mass-selected galaxies. However, this is probably not the case here for the following reason. \citet{2015MNRAS.451.4029K} compared the stellar mass function (SMF) at $z=0$ in 14 various SAMs (including \TCB{a model similar to} B06 and an earlier version of H15 by \citealt{2013MNRAS.431.3373H}, H13). They studied whether SMF variations could be due to different initial mass functions (IMF) assumed in the models (B06 assumes a \citealt{1983ApJ...272...54K} IMF while H15 and H13 use a \citealt{2003PASP..115..763C} IMF). They transformed the stellar masses of galaxies using Chabrier IMF for all the models \TCB{(using the correction from \citealt{2013MNRAS.435...87M})} and showed that the scatter in SMF is only slightly changed by this transformation. Therefore, the specific IMFs of B06 and H15 or H13 are probably not the reason for the different lensing signals in our data. For our galaxy sample, we show the variations in the stellar mass distribution of galaxies between different SAMs (Fig.~\ref{fig:SMFratio}). We find that although both SMFs of \TCB{the Durham models} and G11 differ from that of H15, the GGL and G3L predictions of G11 and H15 are quite consistent, whereas there is a significant difference between the predictions of \TCB{the Durham models} and H15. This makes it unlikely that the discrepant predictions by \TCB{the Durham models} can be attributed to the somewhat different distribution of stellar masses.

{\renewcommand{\arraystretch}{1.3}
\begin{table*}
\caption{Values represent the mean satellite fraction and the mean halo mass over 64 simulated field for the high-$z$ and low-$z$ samples. The standard error of these mean values varies between 0.001 and 0.004 for the satellite fractions and between 0.1 and 0.4 for the halo masses. }
\resizebox{0.47\textwidth}{!}
{
\begin{tabular}{|c|cccc|cccc|}
\hline
\multirow{2}{*}{ high-$z$}   & \multicolumn{4}{c|}{Satellite fraction}                 & \multicolumn{4}{c|}{Halo mass [$10^{13} \Msolar /h$]}\\         
    &  G11              & H15              &   B06   & \TCB{L12}         &  G11            &  H15           & B06   & \TCB{L12}            \\ \cline{1-9}
sm1 &  0.28   &  0.29  &  0.30 & \TCB{0.28}   &  2.6   &  2.8  &  4.2  & \TCB{4.7}   \\
sm2 &  0.33   &  0.32  &  0.34 & \TCB{0.35}    &  3.1   &  3.3  &  4.6  & \TCB{5.2}  \\
sm3 &  0.36   &  0.33  &  0.35 & \TCB{0.37}    &  3.6   &  3.8  &  5.3  & \TCB{5.8}  \\
sm4 &  0.35   &  0.32  &  0.36 & \TCB{0.35}    &  4.2   &  4.6  &  6.1  & \TCB{6.4}  \\
sm5 &  0.32   &  0.30  &  0.34 & \TCB{0.30}    &  4.9   &  5.4  &  6.7  & \TCB{6.9}  \\
sm6 &  0.28   &  0.27  &  0.31 & \TCB{0.25}    &  6.4   &  6.3  &  6.9  & \TCB{7.2}  \\ \cline{1-9}
\end{tabular}}
\hspace{1cm}
\resizebox{0.47\textwidth}{!}
{
\begin{tabular}{|c|cccc|cccc|}
\hline
\multirow{2}{*}{low-$z$}    & \multicolumn{4}{c|}{Satellite fraction}              & \multicolumn{4}{c|}{Halo mass [$10^{13} \Msolar /h$]}\\         
    &  G11            & H15             & B06      & \TCB{L12}        &  G11             & H15             & B06     & \TCB{L12}        \\ \cline{1-9}
sm1 & 0.38  & 0.34  & 0.38  & \TCB{0.36}  &  3.8    &  3.8   & 5.6  &  \TCB{6.0} \\
sm2 & 0.41  & 0.35  & 0.39  & \TCB{0.39}  &  4.2    &  4.2   & 6.0  &  \TCB{6.6} \\
sm3 & 0.40  & 0.35  & 0.37  & \TCB{0.39}  &  4.7    &  4.8   & 6.9  & \TCB{7.4} \\
sm4 & 0.37  & 0.33  & 0.37  & \TCB{0.37}  &  5.3    &  5.5   & 7.8  &  \TCB{8.1}  \\
sm5 & 0.33  & 0.31  & 0.35  & \TCB{0.32}  &  5.9    &  6.4   & 8.5  & \TCB{8.6} \\
sm6 & 0.29  & 0.28  & 0.32  & \TCB{0.26}  &  7.4    &  7.7   & 8.7  & \TCB{9.2} \\ \cline{1-9}
\end{tabular}}
\label{tab:SatelliteFraction}
\end{table*}
}

The discrepancies in model prediction of the lensing signals indicate model variations in the galaxy-matter correlations. It reflects the variations in the way galaxies are distributed among the dark matter halos. This argument is in agreement with the results presented in \citet{2009MNRAS.400.1527K} and \citet{2012A&A...547A..77S} who attributed this trend in B06 to the generation of too many satellite galaxies in massive halos. Indeed, the mean halo masses are higher in \TCB{the Durham models} than in \TCB{the Garching models} for all stellar masses but sm6 and the satellite fraction is somewhat higher for \TCB{the Durham models} (Table~\ref{tab:SatelliteFraction}). One main general difference between the \TCB{Durham and Garching} models is the definition of independent halos and the way descendants of the halos are identified in the merger trees. These differences have an impact on the treatment of some physical processes such as mergers which, in turn, influence the abundance of satellites in halos. Using a halo model description, \citet{2005IAUS..225..243W} 
showed that the galaxy-matter power spectrum, and hence the GGL signal, increases in amplitude when the mean number of galaxies inside halos of a specific mass scale is increased. Similarly, the galaxy-galaxy-matter bispectrum, hence G3L, increases in amplitude if the the number of galaxy pairs is increased for some mass scale. Therefore, an over-production of satellite galaxies in massive halos can explain the relatively high signal of GGL and G3L in \TCB{the Durham models}. This interpretation is supported by the higher mean mass of parent halos of galaxies in \TCB{the Durham models} compared to \TCB{the Garching models} as shown in Table~\ref{tab:SatelliteFraction}.

\newpage

\begin{acknowledgements}
We thank Hendrik Hildebrandt, Douglas Applegate, and Reiko Nakajima for useful discussions. \TCB{We also thank Violeta Gonzalez-Perez and the anonymous referee for fruitful comments.} Hananeh Saghiha gratefully acknowledges financial support of the Deutsche Forschungsgemeinschaft through the project SI 1769/1-1.
Stefan Hilbert acknowledges support by the DFG cluster of excellence \lq{}Origin and Structure of the Universe\rq{} (\href{http://www.universe-cluster.de}{\texttt{www.universe-cluster.de}}).  
\end{acknowledgements}

\bibliographystyle{aa}
\bibliography{BibFiles}
\end{document}